# Wavelet based Watermarking approach in the Compressive Sensing Scenario

Jelena Musić, Ivan Knežević and Edis Franca

*Abstract*— **Due to the wide distribution and usage of digital media, an important issue is protection of the digital content. There is a number of algorithms and techniques developed for the digital watermarking. In this paper, the invisible image watermark procedure is considered. Watermark is created as a pseudo random sequence, embedded in the certain region of the image, obtained using Haar wavelet decomposition. Generally, the watermarking procedure should be robust to the various attacks-filtering, noise etc. Here we assume the Compressive sensing scenario as a new signal processing technique that may influence the robustness. The focus of this paper was the possibility of the watermark detection under Compressive Sensing attack with different number of available image coefficients. The quality of the reconstructed images has been evaluated using Peak Signal to Noise Ratio (PSNR). The theory is supported with experimental results.**

I. INTRODUCTION

Intensive growth of digital communication and technologies causes using digital media in everyday life. However, illegal usage and reproduction of digital content have also intensively growth in recent years. Therefore, there is an intensive development of the methods for digital content protection. One method used for protection of digital media is digital watermarking [1]-[4]. Various signals can be protected using watermarking techniques: audio and video signals, images, etc. Watermark is signal which can be embedded into the original signal and it can be visible or, which is more common, invisible in the original signal. Watermark can be one dimensional signal or can be in the form of logo image. Different domains for watermark embedding can be used: DCT domain [5], [6], DWT domain [7]-[9], time-frequency or space/spatial-frequency transforms [10],[11].

In this paper we will observe image watermarking procedure, where the watermark is embedded into the coefficients of the Haar wavelet transform. Two-dimensional Haar wavelet transform is done in two steps, and embedding region is chosen after the second step. Digital data can be subject of various attacks. Attacks can be intentional with the aim to destroy watermark, or different signal processing techniques may have an impact on watermark detectability. Here, we will observe Compressive Sensing as attack on watermarked image, which is nowadays a popular scenario in watermarking approaches [12]-[16].

Compressive Sensing (CS) is a new approach to the signal sampling [2]. Using different types of optimization algorithms, CS allows signal reconstruction when only small number of samples is available. This reconstruction is possible under certain conditions. Namely, signal should be sparse in certain domain, and samples should be acquired randomly. Two dimensional signals are generally not sparse, in a strict sense. Therefore, in the reconstruction of 2D signals, the gradient is used in the optimization programs. Various algorithms for the reconstruction of the 1D and 2D signals exist: $l_1$- and $l_2$-norm minimization, Total Variation *(TV)* minimization.

The paper is structure as follows: After the Introduction, the theoretical basis related to the watermarking is given in Section II, including the considered watermarking algorithm. In Section III the basic theoretical concepts of the CS and reconstruction methods are given. The experimental results and detection after the CS attack are given in Section IV. Conclusion is given in the Section V.

II. DIGITAL WATERMARKING

Digital watermarking is a technique for multimedia data protection. There is a growing need for multimedia data protection: against illegal copying, distribution and misuse of digital content. Watermarking provides copyright protection, monitoring copies of the content, tamper resistance, authentication and annotation. Watermark is secret digital signal which can be *1D* random sequence, or can be logo image inserted into the multimedia signal. Embedding can be done in time domain or into some of transform domains and work both, for 1D and 2D signals. In this paper we presented image watermark algorithm, where watermark is 1D signal, embedded into the wavelet image coefficients. Let us first give some basics of the watermarking procedures.

Generally the watermark should be imperceptible, secure and robust to all kinds of intentional attacks, such as compressions, noises and geometrical attacks.

*A. Watermark classification*

Watermark can be classified:
- By the human perceptibility to watermark:

Visible or invisible. Visible watermark often represents owner logo image.

- By robustness level:

Fragile watermarking is watermark implementation in the high frequency part of the spectrum, which can be easily removed by high frequency filter.

Semi-fragile watermarking provides resistance to some signal processing techniques.

Robust watermarking is watermark that can't be removed without significant degradation of the signal.

- By transform domain watermark implementation

is performed. Watermark can be inserted into the multimedia signals in time domain or into some of transform domains such as discrete wavelet transform (*DWT*), discrete cosine transform (*DCT*) or discrete Fourier transform (*DFT*).

### B. Watermark embedding

Once the coefficients are chosen, watermark signal can be implemented by additive (1) or multiplication method (2):

$$I_\omega = I_i + \alpha\omega \quad (1)$$

$$I_\omega = I_i + \alpha\omega|I_i| \quad (2)$$

where $I_\omega$ are watermarked coefficients, $I_i$ are coefficients of the original signal, $\omega$ is the watermark and α is the parameter that controls watermark strength (visibility).

### C. Watermark detection

According to necessary of data required for detection:

- Non-blind watermark detection requires using original unwatermarked image in the process.
- Blind detection procedure in which original is not needed.

The watermarking detection method is performed by using standard correlator, which calculates sum of the watermarked coefficients multiplied with the coefficient of watermark, and sum of the watermarked coefficients multiplied with the coefficient of wrong key. The sum is calculated by using following equation :

$$D = \sum w_i DWT_{w_i} \quad (3)$$

$w$ – right key (watermark), *DWT* image coefficients.

The detector response for the right key - watermark should be higher than detector response for any other key - wrong key, that is generated in the same way as the watermark (i.e. generated as a pseudo random sequence):

$$\sum DWT_{w_i} w > \sum DWT_{w_i} wrong \;, \quad (4)$$

Where $DCT_w$ denotes the watermarked coefficients, $w$ is the watermark and *wrong* is the wrong key.

### D. Proposed watermark algorithm

Image representation in some of the transform domains in which image has sparse representation, such as discrete wavelet transform (*DWT*), discrete cosine transform (*DCT*) or discrete Fourier transform (*DFT*), allows good overview of the spectrum and simplifies selection of the area in which watermark is going to be embedded. Wavelet analysis for certain classes of signals and images provides more precise information about signal data than other signal analysis techniques. Wavelet is often used for image compression, feature extraction, signal denoising, data compression, and time-series analysis. Wavelets are functions created by scaling and translation of the base function in the time domain. In this paper 'Haar' wavelet transformation is used for the watermark embedding.

*DWT 'Haar'* transformation, separates an image into a lower resolution approximation image (*LL*), horizontal (*HL*), vertical (*LH*) and diagonal (*HH*) detail components in each decompositionstep. After the obtaining *LL* image approximation in the first step of the Haar wavelet transform, the second step is performed on this region. Fig.1 shows regions of the first and second level decomposition. *DWT* representation allows us to implement higher energy watermarks in region that is less sensitive (such as the high resolution detail bands: *LH, HL, HH*). That increases the robustness of our watermark. After the second step, the *HL* band is chosen to for watermark embedding. Watermark is created as a pseudo random sequence. It is embedded into the *HL* wavelet coefficients, using additive embedding procedure, as described with relation (1).

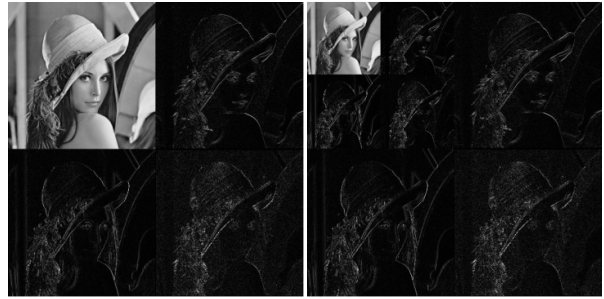

Figure 1. First and second Level Haar image decomposition

The detection is done using standard correlation detector. The detector response for any wrong trial (sequence created in the same manner as watermark) should not be greater than response to the right one. In order to detect watermark from the image first few steps of the embedding procedure are repeated. Once the coefficients of the *HL* band are taken, two sums are being calculated due to equation (4). The procedure is later repeated for the CS reconstructed image.

### III. COMPRESSIVE SENSING

In this section we will summarize a basic concepts of the CS and Total variation minimization. Let us consider a discrete–time signal *x* of length *N*. Any signal can be represented in terms of basis vectors as follows:

$$x = \sum_{i=1}^{N} b_i \psi_i = \psi b \qquad (5)$$

where $b_i$ is weighting coefficient, $\Psi_i$ is basis vector, $\Psi$ denotes $NxN$ transform matrix whose columns are basis vectors and $b$ is the equivalent of the signal in $\Psi$ domain. Reconstruction requires incoherent measurement procedure, i.e. measurement matrix $f$ should be incoherent with the transform matrix $\Psi$ to be successful.

Measurements are stored in vector $v$:

$$\begin{aligned} v_{Mx1} &= \phi_{MxN} x_{Nx1} \\ v &= \phi x = \phi \psi b = Ab \end{aligned} \qquad (6)$$

$\phi$ is measurement matrix and $A$ is CS matrix.

Reconstructed signal can be obtained by using certain optimization algorithm, which can be based on different norms minimization. Optimization technique is based on $l_1$– minimization. The optimization problem is commonly used and defined as:

$$b = \min \|b\|_{l_1} \text{ subject to } v = Ab. \qquad (7)$$

where $b$ is solution of the minimization problem and $\|b\|_{l_1} = \sum_{i=1}^{N}|b_i|$ is $l_1$– norm of vector $b_i$.

Often used method in CS image reconstruction is not $l_1$– minimization but TV minimization, because image is not sparse in the transform domain what is required for the $l_1$– minimization, is more suitable for images. The fact that image gradient is sparse in the transform domain is used in TV minimization and this method is widely used for image denoising. In described method for CS image reconstruction, total number of measurements can be separated in two groups: $v_1$ which are taken from the low frequency $DCT$ image coefficients and $v_2$ middle to high random selected coefficients. Changing the number of those two parameters the quality of the image reconstruction is changing. The quality of the reconstructed images has been evaluated using Peak Signal to Noise Ratio (PSNR). In our reconstruction, if the number of measurements of the total image coefficients is under 25% reconstructed, image is a bit blur.

TV minimization can be defined as:

$$\min_b TV(b) \text{ subject to } v = Ab. \qquad (8)$$

The TV of the signal $b$ represents a sum of the magnitudes of descrete gradients at each point and can be defined as

$$TV(b) = \sum_{i,j} \|D_{i,j} b_{l_2}\|. \qquad (9)$$

$i, j$ is donoted as $D_{i,j}$ and using relation:

$$D_{i,j}b = \begin{bmatrix} b(i+1,j) - b(i,j) \\ b(i,j+1) - b(i,j) \end{bmatrix}. \qquad (10)$$

The total variation functional is a "sum of norms".

$$TV(b) = \sum_{i,j} \sqrt{(b_{i+1,j} - b_{i,j})^2 + (b_{i,j+1} - b_{i,j})^2}. \qquad (11)$$

In our case TV minimization is method used in CS reconstruction of image watermarked by procedure described in Section II.

Reconstruction algorithm required setting up the number of the low frequency $DCT_{cs}$ coefficients, and number of randomly selected coefficients. After the successful reconstruction with no visible image degradation is obtained, the watermark detection is performed using standard correlator.

$$\sum DWT_{csw_i} w > \sum DWT_{csw_i} wrong \qquad (12)$$

Where $DWT_{csw}$ represents watermarked CS coefficients, $w$ is the watermark and *wrong* is the wrong key.

IV. EXPERIMENTAL RESULTS

In our watermarking algorithm wavelet decomposition is preformed on image 'cameraman.tif' size 256x256 on the way that is described in Section II. Watermark, generated as pseudo random sequence vector, is added to the HL coefficients obtained in the second decomposition step, according to the (1). Strength of the watermark is controlled by parameter which multiply generated pseudo random sequence. We set it on 7. Then, we apply inverse DWT to obtain the watermarked image. Fig. 2.a shows the original and watermarked image. After the watermark embedding, we test its robustness to the CS attack. Table below shows PSNR of images reconstructed considering different values of $v_1$ low frequency $DCT$ coefficients and $v_2$ randomly chosen DCT coefficients. In our shown experimental results, it is clear that minimum percent of measurements required for fine reconstruction is 25.9%, below this number, image is blurred and PSNR is low. Image reconstruction result, with just 26% of measurements where $v_1=0$ and $v_2=17000$ is shown on the Fig. 2. b).Taking less number of coefficients will not degrade watermark detection, but will degrade image quality (PSNR of reconstructed image will be below tolerable 30dB). Detection is performed with standard correlator and in our result there is no watermark damage. Fig. 3 shows results of watermark detection, of the original watermarked and CS reconstructed watermarked image.

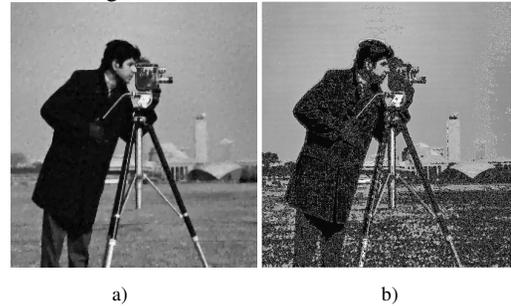

a)        b)

Figure 2. a) Watermarked image in PSNR=46.19, b) CS reconstructed image using 25.9% coefficients in total and PSNR 30.47dB

Blue line represents left sum while red one is sum of the wrong key and watermark coefficients product. The gap between those two lines is big, so the watermark is strong and not damaged. Table 1 provides results in terms of PSNR for various numbers of measurements. We focused on the minimum number of measurements required for good reconstruction (PSNR above 30 dB). It is clear that more low frequency coefficients are taken the better reconstruction quality will be.

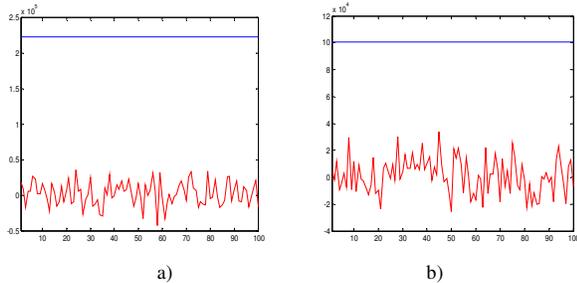

Figure 3.  The results of watermark detection: a) Watermarked image, b) CS attacked image detection using 25.9% coefficients in total

TABLE I.  PSNR OF IMAGES RECONSTRUCTED CONSIDERING DIFFERENT VALUES OF COEFFICIENTS

| Coeff. $K1$ | Coeff. $K2$ | Total % of coeff. | PSNR (db) | Detection |
|---|---|---|---|---|
| **2000** | 12000 | 21% | 29.17 | Succeeded |
| **2000** | 15000 | 25,9% | 30.28 | Succeeded |
| **0** | 17000 | 25,9% | 30,04 | Succeeded |
| **1000** | 17000 | 27,45% | 30,47 | Succeeded |
| **5000** | 15000 | 30,1% | 31,64 | Succeeded |
| **0** | 30000 | 45,77% | 34.75 | Succeeded |

## V. CONCLUSION

In this paper the ability of watermark detection, after CS image reconstruction is analyzed. Watermark, created as a pseudo random sequence, is embedded in the certain region of the image, obtained using second step Haar wavelet decomposition. Image is reconstructed by using different number DCT coefficients as CS measurements. It is shown that reconstructed image has good quality with 25.9 % of total coefficients measurement and that watermark detection succeeded. Watermark detection is successful even for the less number of used measurement, but in such cases image quality is visually degraded.

## VI. ACKNOWLEDGEMENT

The authors are thankful to Professors and assistants within the Laboratory for Multimedia Signals and Systems, at the University of Montenegro, for providing the ideas, codes, literature and results developed for the project CS-ICT (funded by the Montenegrin Ministry of Science).